\documentclass[12pt]{spieman}  
\usepackage{amsmath,amsfonts,amssymb}
\usepackage{graphicx}
\usepackage{setspace}
\usepackage{tocloft}
\usepackage{siunitx}
\usepackage{caption}
\usepackage{lineno}
\usepackage{booktabs, makecell}
\usepackage{soul}
\usepackage{color}
\soulregister\cite7
\soulregister\ref7
\usepackage{fancyhdr}
\usepackage[margin=2.5cm,top=2.5cm,headheight=16pt,headsep=0.50in,heightrounded]{geometry}

\fancypagestyle{FirstPage}{
\lhead{Preprint version. Manuscript submitted to the Journal of Medical Imaging Nov 20, 2019}
\setlength{\headheight}{20pt}
}

\title{DirectPET: Full Size Neural Network PET Reconstruction from Sinogram Data}

\author[a,b,*]{William Whiteley}
\author[b]{Wing K. Luk}
\author[a]{Jens Gregor}

\affil[a]{The University of Tennessee, EECS, 1520 Middle Drive, Knoxville, TN 37996, USA}
\affil[b]{Siemens Medical Solutions USA Inc., 810 Innovation Drive, Knoxville, TN 37932, USA}

\cftpagenumbersoff{figure}
\cftpagenumbersoff{table} 
\begin{document}
\maketitle
\thispagestyle{FirstPage}

\begin{abstract}

\paragraph{Purpose}

Neural network image reconstruction directly from measurement data is a relatively new field of research, that until now has been limited to producing small single-slice images (e.g., 1x128x128). This paper proposes a novel and more efficient network design for Positron Emission Tomography called DirectPET which is capable of reconstructing multi-slice image volumes (i.e., 16x400x400) from sinograms.

\paragraph{Approach}

Large-scale direct neural network reconstruction is accomplished by addressing the associated memory space challenge through the introduction of a specially designed Radon inversion layer. Using patient data, we compare the proposed method to the benchmark Ordered Subsets Expectation Maximization (OSEM) algorithm using signal-to-noise ratio, bias, mean absolute error and structural similarity measures. In addition, line profiles and full-width half-maximum measurements are provided for a sample of lesions.

\paragraph{Results}

DirectPET is shown capable of producing images that are quantitatively and qualitatively similar to the OSEM target images in a fraction of the time. We also report on an experiment where DirectPET is trained to map low count raw data to normal count target images demonstrating the method's ability to maintain image quality under a low dose scenario.

\paragraph{Conclusion}

The ability of DirectPET to quickly reconstruct high-quality, multi-slice image volumes suggests potential clinical viability of the method. However, design parameters and performance boundaries need to be fully established before adoption can be considered.

\end{abstract}


{\noindent \footnotesize\textbf{*}William Whiteley,  \linkable{william.whiteley@siemens-healthineers.com} }

\begin{spacing}{2}   
\setlength{\abovedisplayskip}{3pt}
\setlength{\belowdisplayskip}{3pt}

\section{Introduction}
\label{sect:intro}  

Reconstructing a medical image by approximating a solution to the so-called ill-posed inverse problem typically falls into one of three broad categories of reconstruction methods: analytical, iterative and, more recently, deep learning. While conventional analytical and iterative methods are far more studied, understood and deployed, the recent effectiveness of deep learning in a variety of domains has raised the question whether neural networks are an effective means to directly solve the inverse imaging problem. In this article, we explore an answer to that question for Positron Emission Tomography (PET) with the development of DirectPET, a deep neural network capable of reconstructing a multi-slice image volume directly from Radon encoded measurement data. We analyze the quality of DirectPET reconstructions both qualitatively and quantitatively by comparing against the standard clinical reconstruction benchmark of Ordered Subsets Expectation Maximization plus Point Spread Function (OSEM+PSF)\cite{hudson:97, Rapisarda_2010}. Additionally, we explore the benefits and limitations inherent in direct neural network reconstruction.  

As a precondition, it is reasonable to ask whether medical image reconstruction is an appropriate application for a neural network. The answer to this question is found in the understanding that feed-forward neural networks have been proven to be general approximators of continuous functions with bounded input under the Universal Approximation Theorem \cite{Hornik:91,csaji:01,zhou:17}. The nature of PET imaging makes it such a problem with the implication that a solution can be approximated by a neural network. This leads us to believe that the study of direct neural network reconstruction is a worthy pursuit.

We acknowledge that the notion of direct neural network reconstruction may be somewhat controversial. The primary criticism, which the authors freely admit is reasonable, is that it foregoes decades of imaging physics research as well as the careful development of realistic statistical models to approximate the system matrix, not to mention corrections for scatter and randoms. Instead of utilizing these handcrafted approximations, data driven reconstruction  solves the inverse problem by directly learning a mapping between measurement data and images from a large number of examples (targets) and in turn encodes this mapping into millions or billions of network parameters. The disadvantage with the method is its black box nature and the current inability to understand and explain the reasoning behind a given set of trained parameters for networks of any significant size or complexity. 

Speaking in favor of direct neural network reconstruction, on the other hand, are distinct and quantifiable benefits not found with conventional methods. First and foremost, we will show that direct neural network reconstruction provides good image quality with very high computational efficiency once training has been completed. Specifically, we show that the DirectPET network can produce a multi-slice image volume comparable to OSEM+PSF in a fraction of the amount of time needed by the model-based iterative method. Another benefit is the adaptability and flexibility that deep learning methods provide in that the output can be tuned to exhibit specific desirable image characteristics by providing training targets with those same characteristics. In particular, data driven reconstruction methods can produce high quality images, if data sets containing high quality image targets are available to train the neural network. We demonstrate this ability by showing that DirectPET can be trained to learn a mapping from low count sinograms to high count images. Subsequent reconstruction produces images of a quality that is superior to those produced by OSEM-PSF.

The proposed DirectPET network advances the applicability of direct neural network reconstruction. AUTOMAP\cite{Zhu:2018} and DeepPET\cite{HAGGSTROM:2019} have only been shown to produce single-slice 1x128x128 images. AUTOMAP  has specifically been critiqued\cite{Wang:18} for the image size being limited by its large memory space requirement. For direct methods to be of practical relevance, they must be able to produce larger image sizes as commonly used in clinical practice. DirectPET was designed for efficiency, and we demonstrate single forward pass simultaneous reconstruction of multi-slice 16x400x400 image volumes. When batch operations are employed, DirectPET can not only reconstruct an entire full-size 400x400x400 whole-body PET study, but does so in a little more than one second.

In this paper, the general advantages of direct neural network reconstruction methods and the specific advancements of the proposed DirectPET network are explored with the specific contributions being as follows:

\begin{enumerate}
  \item\textit{A novel direct reconstruction neural network design}: A three segment architecture capable of full-size multi-slice reconstruction along with a quantitative and qualitative analysis compared to the conventional reconstruction benchmark of OSEM+PSF.  
  \item\textit{Effective direct reconstruction neural network training techniques}: We propose specific techniques (loss function, hyper-parameter selection and learning rate management) not utilized in previous direct reconstruction methods to achieve efficient learning and high image quality overcoming the often blurry images produced by neural networks using simple L1 or L2 loss functions.  
  \item\textit{A path to superior image quality}: We demonstrate that the image quality depends less on the raw data and more on the target images used for training.  
  \item\textit{Challenges and a path to overcoming them}: We discuss current challenges and limitations associated with direct neural network reconstruction and propose a future path to reliably surmounting these obstacles. 
\end{enumerate}

\section{Related Work}
The terms deep learning and image reconstruction are often used in conjunction to describe a significant amount of recent research\cite{Wang:18} that falls into one of three categories: 1) combination of deep learning with a conventional analytical or statistical method; 2) use of a neural network as a nonlinear post reconstruction  filter for denoising and mitigating artifacts; and less commonly 3) use of a neural network to generate an image directly from raw data. This last category of direct neural network reconstruction, which forms the largest departure from conventional methods, is the focus of our work.

Early research was based on networks of fully connected multilayer perceptrons \cite{Floyd:91,Bevilacqua:00,Paschalis:04,Argyrou:12} that yielded promising results, but only for simple low resolution reconstructions. More recent efforts have capitalized on the growth of computational resources, especially in the area of Graphical Processing Units (GPUs), which has led to deep networks capable of direct reconstruction. The AUTOMAP network \cite{Zhu:2018} is a recent example that utilizes multiple fully connected layers followed by a sparse convolutional encoder-decoder to learn a mapping manifold from measurement space to image space.  AUTOMAP is capable of learning a general solution to the reconstruction inverse problem. However, the generality is achieved by learning an excessively high number of parameters which limits its application to fairly small single-slice images (e.g., 1x128x128). DeepPET\cite{HAGGSTROM:2019} is another example of direct neural network reconstruction that utilizes an encoder-decoder architecture, but forgoes any fully connected layers. Instead, this network utilizes convolutional layers to encode the sinogram input (1x288x269) into a higher dimensional feature vector representation (1024x18x17) which is then decoded by convolutional layers to produce yet another small single-slice image (e.g., 1x128x128). While both of these novel methods embody significant advancements in direct neural network reconstruction, there are several noteworthy differences to the DirectPET network presented here. As mentioned above, DirectPET is capable of producing multi-slice image volumes (e.g., 16x400x400). We furthermore train and validate DirectPET on actual raw PET data taken from patient scans as opposed to using simulated data. Also not done previously, we include the attenuation maps as input to the neural network. Finally, DeepPET was shown to become unstable at low count densities generating erroneous images \cite{HAGGSTROM:2019}, a problem not encountered for DirectPET which exhibits consistent performance across all count densities.

A currently more common application of deep learning in the image formation process is combining a neural network with conventional reconstruction. One method is using an image-to-image neural network to apply an image-space operator to enhance the post reconstruction output. While the term "reconstruction" has been attached to some of these methods, the neural network is not directly involved in solving the inverse imaging problem, and is more accurately described as a post reconstruction filter. These learned filters have produced improvements compared to conventional handcrafted alternatives. In PET imaging, these image-space methods are often applied to low dose images to produce normal dose equivalents \cite{Gong:18,Kaplan:2018,Xu:2017} utilizing U-Net \cite{Ronneberger:15} or ResNet \cite{He:15} style networks. Similarly, these methods are demonstrated to work for X-ray CT on low dose image restoration \cite{Kang:18} and limited angle\cite{Han:18} applications. Jiao \textit{et al.}\cite{Jiao:2017} increased the reconstruction speed by performing a simple back-projection of PET data and then utilizing a neural network to reduce the typical streaking artifacts. Cui \textit{et al.}\cite{Cui2019} used a novel unsupervised approach to denoise a low count PET image from the same patient's previous high quality image. 

As an alternative to the image-space methods, other efforts have included deep learning elements inside the conventional iterative reconstruction process creating unrolled neural network methods. This often takes the form of using the neural network as a denoising or regularizing operator inside the iterative loop for both PET \cite{Kim:18, Gong:2018:2} and X-ray CT \cite{Chen:2018, Gjesteby:17} reconstruction. Gong \textit{et al.} \cite{Gong:19} replaced the penalty gradient with a neural network in their Expectation Maximization network (EMnet) and Alder \textit{et al.} \cite{Adler:2018} replaced the proximal operators in the primal-dual reconstruction algorithm. While  the image-space and unrolled deep learning methods all demonstrated improvement over conventional handcrafted alternatives, in addition to now containing a black box component in the form of a neural network, they continue to carry all of the disadvantages of conventional reconstruction methods, namely, the complexity of multiple projection and correction components and a high computational cost. By comparison, direct neural network reconstruction methods are relatively simple operators with very high computational efficiency once trained.

\section{Methods}

\subsection{Network Architecture Overview}

\begin{figure*}[h!]
	\centering
	\includegraphics[width=\textwidth]{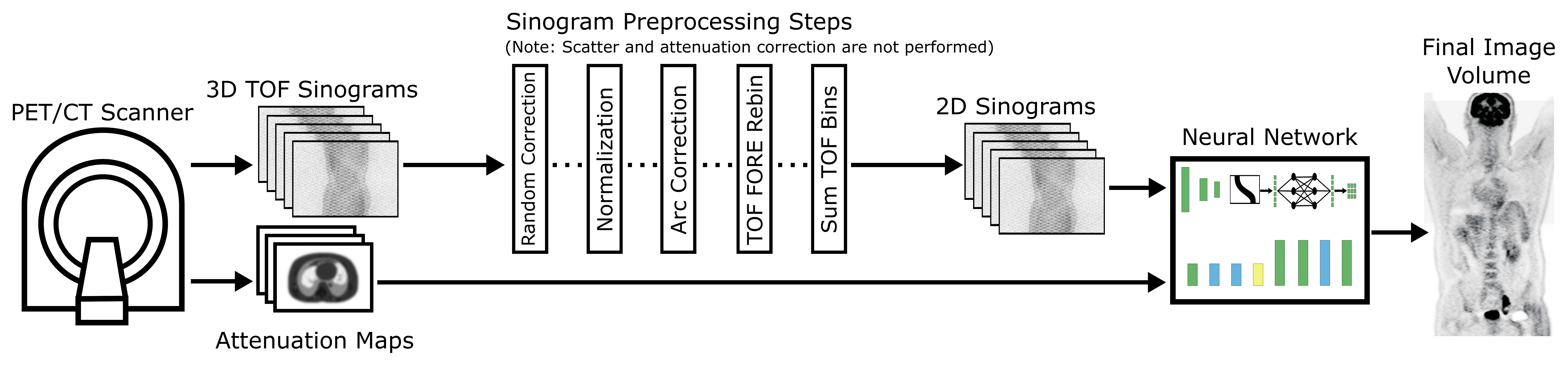}
	\caption{The DirectPET imaging pipeline uses TOF Fourier rebinned PET data and X-ray CT based attenuation maps to generate PET image volumes.}
	\label{pipeline}
\end{figure*}

Figure~\ref{pipeline} illustrates where the proposed DirectPET network fits in the PET imaging pipeline. Time-of-flight (TOF) list-mode data is acquired and histogrammed into sinograms.  Random correction, normalization and arc correction are performed on the raw sinogram data in that order. Scatter and attenuation correction are not applied in the sinogram domain but instead accounted for by the learned parameters in the DirectPET neural network.  Oblique plane sinograms are eliminated by applying Fourier rebinning \cite{Defrise:2005}, and the TOF dimension of the resulting direct plane sinograms is collapsed. The sinogram data could have been transformed to 2D Non-TOF data in a single rebinning step \cite{Li_2015}, but was done in two steps for convenience with readily available software.  X-ray CT data is acquired to create attenuation maps and for later anatomical visualization. A single forward pass through the DirectPET network of the PET/CT data then produces the desired multi-slice image volume.

\begin{figure*}[h!]
	\centering
	\includegraphics[width=\textwidth]{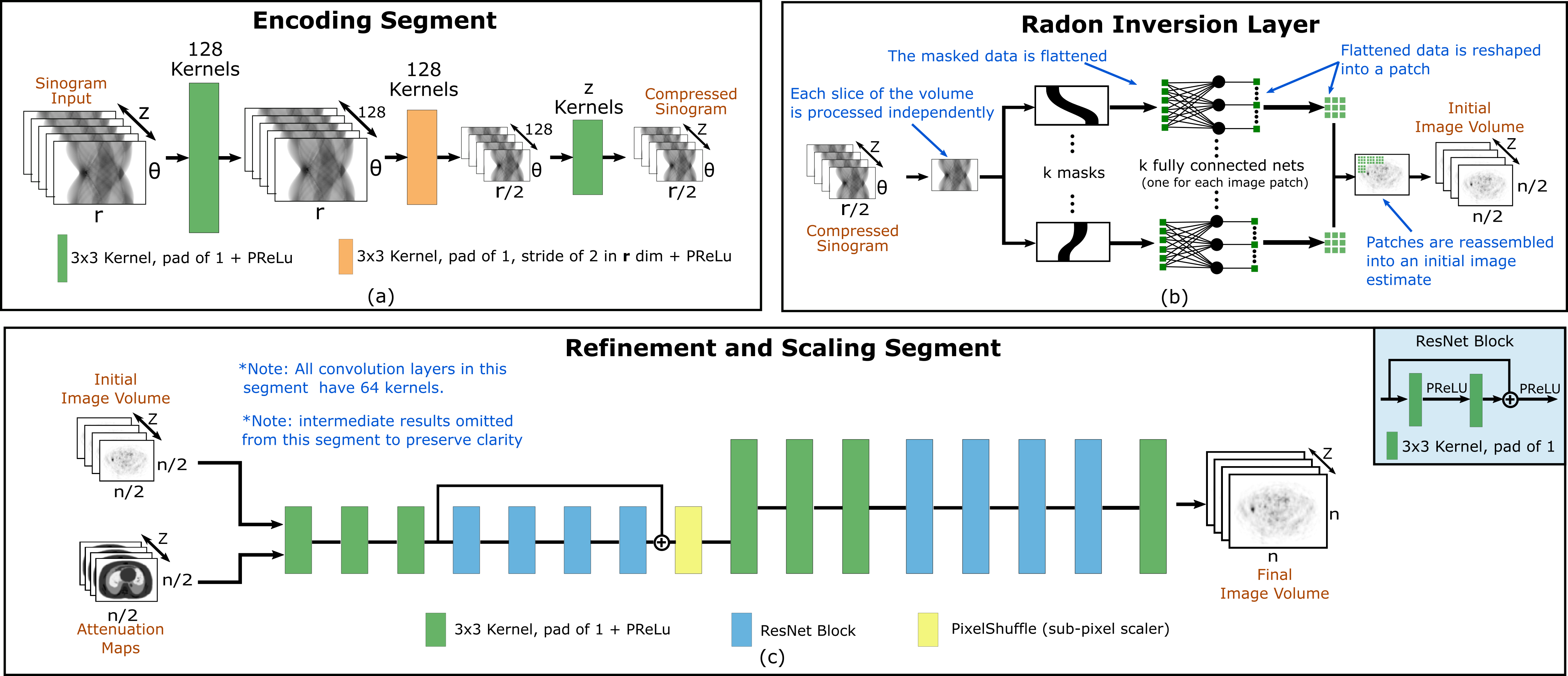}
	\caption{The DirectPET reconstruction neural network consists of three distinct segments each with a specific task: (a) the encoding segment is composed of convolutional layers that compress the sinogram input; (b) the domain transformation segment implements Radon inversion by applying masks to filter the compressed sinogram data into small fully connected networks for each of a number of image patches that are then combined to produce an initial image estimate; and (c) the refinement and scaling segment carries out denoising along with attenuation correction and applies super-resolution techniques to produce a final full-scale image.}
	\label{detailed_network}
\end{figure*}

With reference to Figure~\ref{detailed_network},  DirectPET consists of three distinct segments each designed for a specific purpose. An encoding segment compresses the sinogram data into a lower dimensional space. A domain transformation segment uses specially designed data masking along with small fully connected layers to carry out the Radon inversion needed to convert the compressed sinogram into image-space. Finally, a refinement and scaling segment enhances and upsamples an initial image estimate to produce the final multi-slice image volume. Together, the segments comprise an Encoding, Transformation and Refinement and Scaling (ETRS) architecture \cite{whiteley:193d}. We proceed by describing each segment starting with the domain transformation layer, which is what enables the DirectPET efficiency, followed by the encoder and then the refinement and scaling segment.

\subsubsection{Domain Transformation}

\begin{figure*}[h!]
	\centering
	\includegraphics[width=\textwidth]{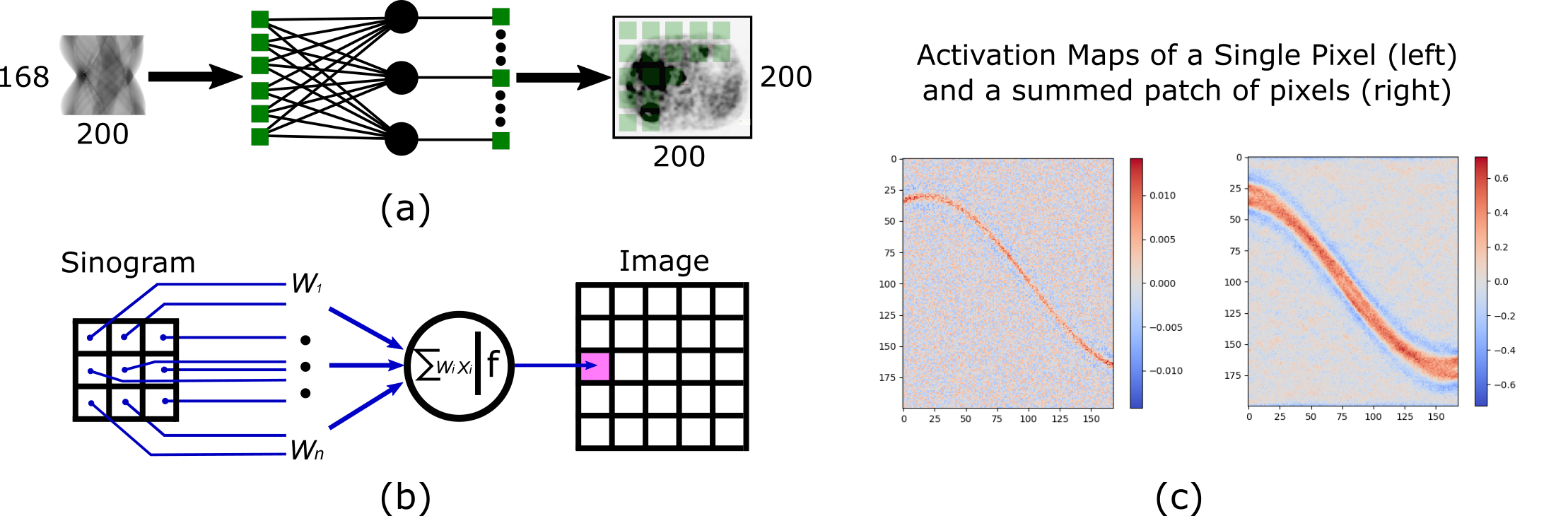}
	\caption{
	  A single, fully-connected layer can be trained to learn the distinctive sinusoidal pattern associated with the Radon transform. 
	}
	\label{intuition}
\end{figure*}

The computational cost of performing the domain transformation from sinogram to image space is a key challenge with direct neural network reconstruction. In previous research \cite{Floyd:91,Bevilacqua:00,Paschalis:04,Argyrou:12,Zhu:2018} this was typically accomplished through the use of one or more fully connected layers where every input element is connected to every output element. This results in a multiplicative scaling of the memory requirements proportional to the size of the input and the number of neurons, i.e., the number of sinogram bins and the number of image voxels. Figure~\ref{intuition}(a,b) illustrates a simple, single layer reconstruction experiment where each bin in a 200x168 sinogram is connected to every pixel in a corresponding 200x200 image. After training to convergence using a natural image data set, examination of the learned activation maps shown in Figure~\ref{intuition}(c) reveals that network learned an approximation to the inverse Radon transform. Along with noting the sinusoidal distribution of the activation weights, the other key observation is that the majority of weights in the activation map are near-zero, meaning they do not contribute to the output image and essentially constitute irrelevant parameters.

With insight from this initial experiment, we designed a more efficient Radon inversion layer to perform the domain transformation. We eliminated network connections that do not contribute to the output image by creating small fully-connected networks for each patch in the output image that are only connected to the relevant subset of sinogram data. The activation maps from the initial simple experiment were used to create a sinogram mask for each patch in the image. Each of these masks is then independently applied to the compressed sinogram, and the surviving bins are fed to an independent fully-connected layer connected to the pixels in the relevant image patch. These patches are then reassembled to create the initial image estimate. When a multi-slice image volume is reconstructed, the transformation is carried out independently for each slice in the stack, and the volume is reassembled at output of the segment. The primary design decision for this segment is selecting the size of the image patch to consider, and is a trade-off between execution speed and memory consumption. Considering that a downsized sinogram (i.e., 168x200) is the input to the Radon inversion layer and a half-scale image estimate (i.e., 1x200x200) is the output, on one end of the spectrum a patch size of a single pixel results in 31,415 fully connected networks (we only address pixels in the field of view). On the other end of the spectrum, if the patch size equals the entire image, only a single fully connected network is required, but this choice requires the maximum 1.055 billion network parameters. We have settled on using 40x40 pixel patches as a baseline, but have empirically found that patches with 30x30 to 50x50 pixels provide a good trade-off between speed and memory consumption. Table \ref{mask_table} shows the parameter count as a function of patch size selection for 168x200 sinograms and 200x200 images. The parameter count for AUTOMAP and a single fully-connected layer are included for comparison.

\begin{table}[h]
\centering
\caption{The selection of patch size is directly related to the required number of parameters in the transform segment and inversely related to the number of masks and associated networks impacting the execution speed.}
\label{mask_table}
\resizebox{\columnwidth}{!}{%
\begin{tabular}{|l|c|c|c|c|c|}
\hline
\multicolumn{1}{|c|}{\textbf{Network}} & \textbf{Patch Size} & \textbf{Input Size} & \textbf{Output Size} & \textbf{Segment Parameters} & \textbf{Number of Masks} \\ \hline
AUTOMAP & na & 200 x 168 & 200 x 200 & 6,545,920,000 & na \\ \hline
Fully-Connected Layer & 200 x 200 & 200 x 168 & 200 x 200 & 1,055,544,000 & 1 \\ \hline
Radon Inversion Layer & 60 x 60 & 200 x 168 & 200 x 200 & 627,224,400 & 16 \\ \hline
Radon Inversion Layer & 40 x 40 & 200 x 168 & 200 x 200 & 382,259,200 & 28 \\ \hline
Radon Inversion Layer & 30 x 30 & 200 x 168 & 200 x 200 & 353,583,000 & 52 \\ \hline
Radon Inversion Layer & 20 x 20 & 200 x 168 & 200 x 200 & 238,370,400 & 88 \\ \hline
Radon Inversion Layer & 10 x 10 & 200 x 168 & 200 x 200 & 209,706,900 & 336 \\ \hline
\end{tabular}
}
\end{table}

\begin{figure*}[h!]
	\centering
	\includegraphics[width=\textwidth]{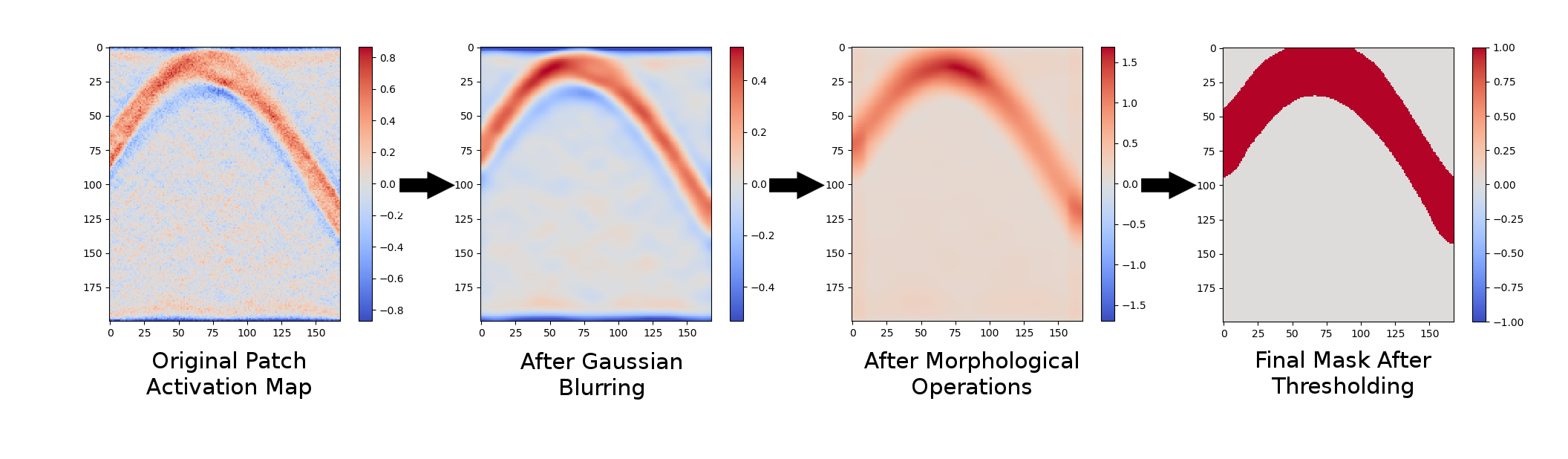}
	\caption{
	  The mask creation process begins with summing the raw pixel activation maps for an image patch and then undergoes a process of smoothing, morphological opening and closing, and thresholding to produce the final mask.  
	}
	\label{masks}
\end{figure*}

Having selected the patch size, the learned activation maps for each pixel in a patch are summed together. However, the raw activation maps are noisy and applying a simple threshold to generate a mask includes sinogram bins that should not contribute to a given image patch. The activation maps are consequently refined using a three step process of Gaussian smoothing to filter high frequency noise, morphological opening and closing operations to remove noise and fill gaps, and thresholding using Li's iterative minimum cross entropy method \cite{LI:93}. Figure~\ref{masks} illustrates the mask refining process. The resulting size of the learned mask can also be tuned by adjusting the size of the Gaussian filter (here $\sigma\!=\!4$) and the morphological structuring element (here a disk of radius 8). A somewhat simpler approach that generates less memory efficient masks, but achieves comparable results, would be to create sinogram masks by forward projecting image patches surrounded by a small buffer\cite{whiteley:2019}.  

\subsubsection{Encoding}
Despite the significant efficiency gains achieved in the domain transformation segment, the original uncompressed sinogram is still too large to process with only modest computational resources. We initially explored simple bilinear scaling and angular and line of response compression/summing, but achieved superior performance allowing a convolutional encoder to learn the optimal compression. The theory and motivation behind this segment is similar to that found in the first half of an autoencoder\cite{Rumelhart:86} where the convolutional kernels extract and forward essential information in each successive layer. Figure~\ref{detailed_network}(a) shows a detailed diagram of the encoding segment illustrating its architecture and chosen hyper-parameters consisting of three convolutional layers each with 128 3x3 kernels and a Parametric Rectified Linear Unit (PReLU)\cite{He:2015} activation function. Spatial down-sampling is accomplished by the second convolutional layer employing a kernel stride of 2 along the $r$ sinogram dimension. While we experimented with the axial dimension (number of slices) of the input between 1 and 32 slices, we ultimately settled on training DirectPET on 16 slices.

\subsubsection{Refinement and Scaling}

The final neural network segment is responsible for taking the initial image estimate plus the corresponding attenuation maps and removing noise and scaling the image to full size. The attenuation maps added at this point in the network provide additional image space anatomical information which significantly boosts the network's image quality. The refinement and scaling tasks draw on significant deep learning research in the areas of denoising\cite{Chunwei:19} and super-resolution\cite{yang:19}. As illustrated in Figure~\ref{detailed_network}(c), the refinement and scaling segment  uses a simple two stage strategy, where each stage contains convolutional layers followed by a series of ResNet\cite{He:15} blocks that include an overall skip connection. This strategy is employed first at half spatial resolution and then again at full resolution after a sub-pixel transaxial scaling by a factor of 2 using the PixelShuffle\cite{Shi:2016} technique. These two sub-segments are followed by a final convolutional layer that outputs the image volume. All layers in this segment use 64 3x3 convolutional kernels and PReLU activation.

\subsection{Neural Network Training}
The DirectPET network was implemented with the PyTorch\cite{paszke:2017} deep learning platform and  executed on single and dual Nvidia Titan RTX GPUs. Training occurred over 1,000 epochs with each epoch having 2,048 samples of sinogram and image target pairs randomly drawn from the training data in mini-batches of 16. The Adam optimizer\cite{Kingma:14} was used with $\beta_1\!=\!0.5$ and $\beta_2\!=\!0.999$, which is similar to traditional stochastic gradient descent but additionally maintains a separate learning rate for each network parameter. In addition to the optimizer, a cyclic learning rate scheduler\cite{Smith:15} was employed that cycles the learning rate between a lower and upper bound with the amplitude of the cycle decaying exponentially over time towards the lower bound. This type of scheduler aids training since the periodic raising of the learning rate provides an opportunity for the network to escape sub-optimal local minimum and traverse saddle points more rapidly.

Based on a triangular wave function, our scheduler was defined as follows for the $k$th iteration:
\begin{subequations}
\begin{gather}
  \eta(k) =
    \Lambda(k) \: (\eta_{max}-\eta_{min}) \: 0.99995^n+\eta_{min} \\
  \Lambda(k) \triangleq
    2 \left\lvert \frac{k}{1000} -
        \left\lfloor \frac{k}{1000} + \frac{1}{2} \right\rfloor
      \right\rvert
\end{gather}
\end{subequations}

To determine appropriate values for the two bounds, an experiment was performed where the learning rate was slowly increased and plotted against the loss. The results of this study led us to designate $\eta_{min}=0.5\times10^{-5}$ and $\eta_{max}=9.0\times10^{-5}$.

The loss function is another primary component of neural network training. Previous research\cite{Zhao:17} dedicated to loss functions for image generation and repair suggested a weighted combination of the element-wise L1 loss, which is an absolute measure, and a multi-scale structural similarity (MS-SSIM) \cite{wang:04} loss, which is a perceptual measure. We extended this idea by eliminating the static weighting factor and instead developed a dynamically balanced scale factor $\alpha$ between these two elements. We also added an additional perceptual feature loss component based on a so-called VGG network \cite{Simonyan:14}. The loss between reconstructed image $\hat{x}$ and target image $x$ was thus made to consist of three terms, namely:

\begin{align}
\mathcal{L}(\hat{x},x) &= \beta \ \text{VGG}(\hat{x},x) + (1-\alpha) \ \text{MAE}(\hat{x},x) + \alpha \ \text{MS-SSIM}(\hat{x},x)
\label{loss_fun}
\end{align}

The VGG loss is based on a convolutional neural network of the same name. Pre-trained on the large ImageNet data set, which contains millions of natural images, each convolutional layer of the VGG network is a general image feature extractor; earlier layers extract fine image details like lines and edges, while deeper layers extract larger semantic features as the image  is spatially down sampled. In our application, the output from the DirectPET network and the target image are independently input to the VGG network. After the input passes through each of the first four layers, the output is saved for comparison. The features from the DirectPET image are then subtracted from the target image features for each of the four VGG layers. The sum of the absolute value differences forms the VGG loss. That is:

\begin{align}
\text{VGG}(\hat{x},x) &= \sum_{\ell=0}^3 |\text{VGG}^\ell(\hat{x})-\text{VGG}^\ell(x)| .
\label{VGG_loss}
\end{align}

Thus accounting for perceptual differences helps the DirectPET network reconstruct images of higher fidelity than otherwise possible.

The MAE loss denotes the Mean Absolute Error between reconstructed image $\hat{x}$ and target image $x$ calculated over all $N$ voxels:

\begin{align}
\text{MAE}(\hat{x},x) &= \frac{1}{N}\sum_{i=0}^{N-1}|\hat{x}_i-x_i| .
\label{l1_loss}
\end{align}

The MS-SSIM loss measures structural similarity between the two images based on luminance (denoted $l_M$), contrast (denoted $c_j$), and structure (denoted $s_j$) components calculated at $M$ scales. More specifically:

\begin{align}
\text{MS-SSIM}(\hat{x},x) &= 1 - l(\hat{x},x) \prod_{j=1}^M c_j(\hat{x},x) s_j(\hat{x},x)
\label{ssim_loss}
\end{align}

where 

\begin{align}
l(\hat{x},x) &= \frac{2\mu_{\hat{x}}\mu_x+C_1}{\mu_{\hat{x}}^2+\mu_{x}^2+C_1} ,   
\nonumber \\
c_j(\hat{x},x) &= \frac{2\sigma_{\hat{x}}\sigma_x+C_2}{\sigma_{\hat{x}}^2+\sigma_{x}^2+C_2} ,
\nonumber \\
s_j(\hat{x},x) &= \frac{\sigma_{\hat{x}x}+C_3}{\sigma_{\hat{x}}\sigma_x+C_3} .
\nonumber  
\end{align}
As usual, $\mu_{\hat{x}}$ and $\mu_x$ denote the two image means, and $\sigma_{\hat{x}}^2$ and $\sigma_{x}^2$ are the corresponding variances while $\sigma_{\hat{x}x}$ is the covariance.
The constants are given by
$C_1\!=\!(K_1 L)^2$, $C_2\!=\!(K_2 L)^2$, and $C_3\!=\!C_2/2$ where $L$ is the dynamic range of values while $K_1\!=\!0.01$ and $K_2\!=\!0.03$ are generally accepted stability constants. 

With respect to the weighting of the VGG loss,
we used $\beta\!=\!0.5$ for all updates. In contrast, 
we used a dynamically calculated value for
$\alpha$ that trades off the
MAE and MS-SSIM losses against one another.
That is:

\begin{align}
   \alpha=\frac{\sum\limits_{j=i}^{i+n-1}\text{MAE}_j}{ \sum\limits_{j=i}^{i+n-1}\text{MAE}_j+\sum\limits_{j=i}^{i+n-1}\text{MS-SSIM}_j}
\end{align}
where $i$ and $j$ are iteration steps and $n$ denotes the width of a running average window.

\section{Experiments and Results}

We now describe details of our training, validation and test data sets, and then evaluate  performance of the DirectPET network in three areas. First, we examine the improvement in reconstruction speed using the proposed method versus conventional iterative and analytical reconstruction methods. Next, we evaluate the quantitative performance of the proposed method on measures of mean absolute error, structural similarity, signal-to-noise ratio (SNR), and bias. We also evaluate two different lesions by comparing line profiles, the full-width half-maximum (FWHM), and zoomed images. Lastly, we  review the image quality by examining patient images from various anatomical regions with varying count levels. 

\subsection{Training and Validation Data}
The data set is derived from 54 whole-body PET studies with a typical acquisition duration of 2-3 minutes per bed for a total of 324 field-of-views (FOVs) or 35,316 individual slices. PET whole-body data sets are particularly challenging because the range of anatomical structures and noise varies widely. Figure~\ref{histo_timing}(a,b) illustrates the count density across all Fourier rebinned sinograms showing slices ranging from 34,612 to 962,568 coincidence counts with a mean value of 218,812 counts. All data was acquired on a Siemens Biograph mCT \cite{Rausch:15} and reconstructed to produce 400x400 image slices using the manufacturer's standard OSEM+PSF TOF reconstruction with 3 iterations and 21 subsets including X-ray CT attenuation correction and a 5x5 Gaussian filter. These conventionally reconstructed images constitute the training targets for the DirectPET network with the goal of producing images of similar quality. At the outset, 14 patients were set aside with 4 going into a validation set used to evaluate the model during training and 10 patients comprising the test set only used to evaluate the final model. Although it is common to normalize input and target data for neural network training, since we are interested in a quantitative reconstruction, the values are not normalized but instead the input sinograms and target images are scaled by a fixed value to create average data values closer to 1, which is conducive to stability during neural network training. With this in mind, sinogram counts were scaled down by a factor of 5 and image voxel Bq/ml values were scaled down by a factor of 400. During analysis the images were scaled back to their normal range to perform comparisons in the original units.

\subsection{Reconstruction Speed}

One of the most pronounced benefits of direct neural network reconstruction is the computational efficiency of a trained network and the resulting speed up in the subsequent image formation process. While a faster reconstruction may not greatly benefit a typical static PET scan, studies where a large number of reconstructions are required such as dynamic or gated studies with many frames or gates will see a significant benefit. In these cases, where conventional reconstruction could take tens of minutes, it would be reduced to tens of seconds. A fast reconstruction would also allow a radiologist reading a study to rerun the reconstruction multiple times with different parameters very quickly if desired. Admittedly this would require the training of a plurality of neural networks to choose from, but we believe this is what will ultimately be required for direct neural network reconstruction similar to the selection of iterations, subsets, filter and scatter correction in conventional reconstruction to produce the most diagnostically desirable image. Additionally, very fast reconstruction could enable entirely new procedures such as interventional PET where a probe is marked with a radioactive tracer, or performing radiation or proton therapy tumor ablation in a single step with real-time patient imaging/positioning and treatment versus the two-step process common today. 

\begin{figure*}[h]
	\centering
	\includegraphics[width=\textwidth]{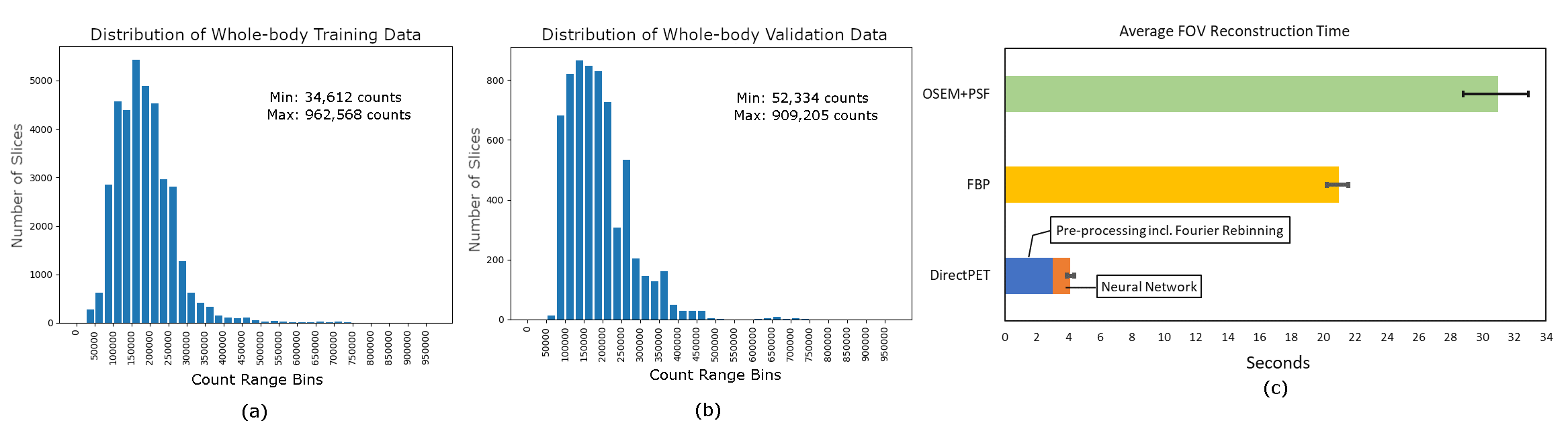}
	\caption{
	  The histograms in (a) and (b) show the relative distribution of slice counts in the Fourier rebinned sinograms for the training and test sets. (c) Shows the reconstruction time of a single FOV for both conventional methods and DirectPET demonstrating 7.2x and 4.9x improvement respectively.
	}
	\label{histo_timing}
\end{figure*}

A comparison of reconstruction speed is shown in Figure~\ref{histo_timing}(c) where all three methods start from the same set of oblique TOF sinograms and end with a final whole-body image volume. The test was run on all 10 whole-body data sets, and the average time shown for each method refers to reconstruction of a single FOV (which makes up 109 image slices).
We followed standard clinical protocols. OSEM+PSF TOF reconstruction was based on 21 subsets and 3 iterations. We used attenuation and scatter correction as well as post-reconstruction smoothing by a 5x5 Gaussian filter. Filtered Back-Projection (FBP) included the same corrections and filtering. Both reconstructions were performed on an HP Z8 G4 workstation with two 10-core Intel Xeon Silver 4114 CPUs running at 2.2 GHz. DirectPET utilized a patch size of 40 x 40 pixels and a batch size of 7. These reconstructions were performed on an HP Z840 workstation with an Intel E5-2630 CPU running at 2.2 GHz and a single Nvidia Titan RTX GPU. 

The results show that reconstruction with the DirectPET image formation pipeline on average takes 4.3 seconds from start to finish. Is is noteworthy that 3 seconds are dedicated to pre-processing and the Fourier rebinning while a mere  1.3 seconds is needed for the forward reconstruction pass of the network. If additional realizations were desired with different reconstruction parameters, the additional DirectPET reconstructions would only require the 1.3 second forward pass of the network. In comparison, OSEM+PSF TOF reconstruction of the same data averaged 31 seconds while FBP averaged 21 seconds, which is 7.2x and 4.9x slower respectively. 

\subsection{Quantitative Image Analysis }

For this section on quantitative performance and the next section focusing on qualitative aspects, two versions of DirectPET were trained. The first version, which will be referred to as DirectPET, was trained using the entire full count PET whole-body training set. The second version, which will be referred to as DirectPET-50, was trained with half of the raw counts removed using list-mode thinning while retaining the full count images as training targets. This second network evaluates the ability of the DirectPET network to produce high quality images from low count input data. For comparison, the  half-count raw data was also reconstructed with OSEM+PSF TOF using the same reconstruction parameters as the full-count data. Those images are referred to as OSEM+PSF-50.

\begin{figure*}[h]
	\centering
	\includegraphics[width=\textwidth]{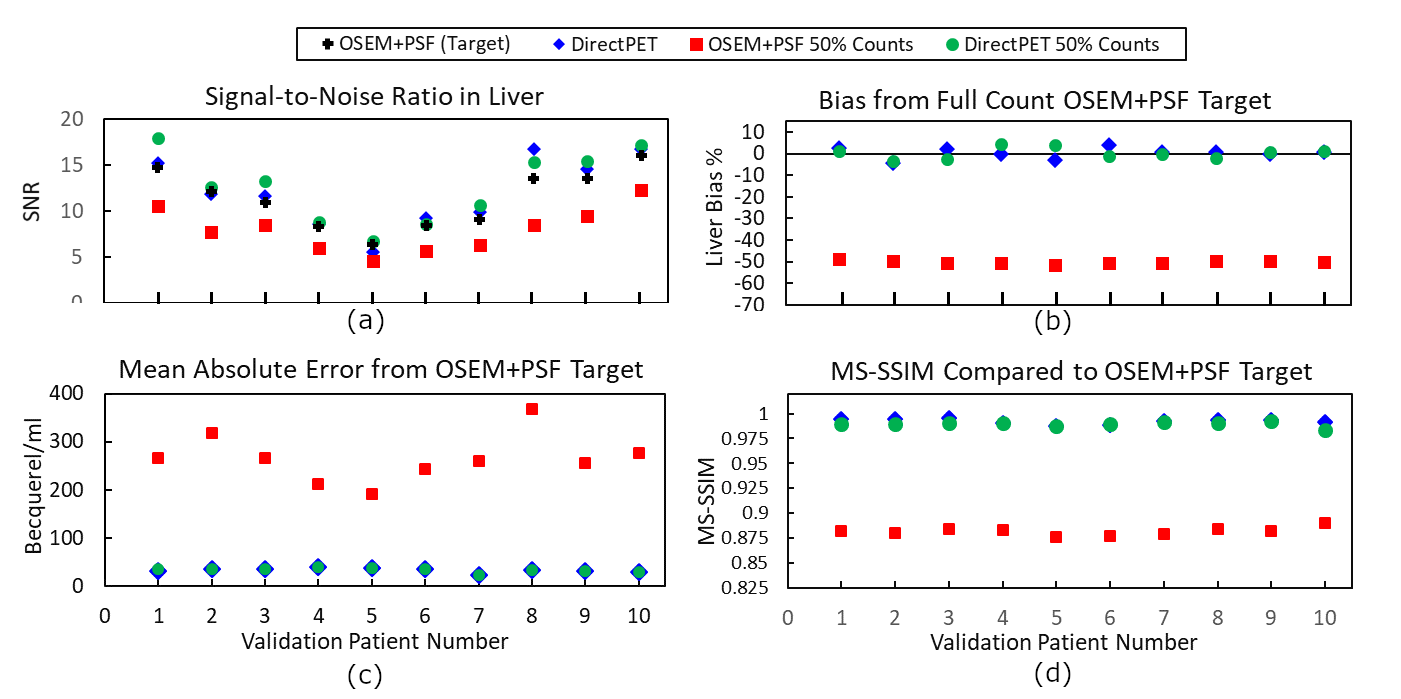}
	\caption{
	  Quantitative measurements of the reference OSEM+PSF reconstructions, DirectPET, DirectPET trained on half-count input data and OSEM+PSF reconstructed with half-count input data show (a) the neural network methods produce images with less noise, (b) introduce a non-systematic bias of less than 5\%, (c) produce low absolute voxel error, and (d) produce images that are perceptually very similar to the target. These plots also show that the neural network trained with only half of the raw counts performed about the same as the network trained on full counts from a quantitative perspective. }
	\label{quant}
\end{figure*}

Figure~\ref{quant}(a) is the measured SNR over an average of 3 VOIs in areas of relative uniform uptake in the liver for each of the 10 patients in the test set. The DirectPET and DirectPET-50 networks both exhibit similar but slightly higher SNR values compared with the target OSEM+PSF. This indicates the neural network produces slightly smoother images with less noise. Smoothing is a common feature of neural networks due to the data driven way they optimize over a large data set. Smoother images are, of course, only acceptable if structural details are preserved along with spatial resolution. This is explored below. As one would predict, OSEM+PSF-50 reconstructions exhibit the lowest SNR indicating a lack of ability to overcome lower count input data. 

Bias measurements, which were calculated relative to the target OSEM+PSF images, indicate if there is an overall mean deviation from the target value. Again these measurements were calculated from three volumes-of-interest (VOIs) in the liver of each patient and averaged. The results shown in Figure~\ref{quant}(b) first indicate there is no global systematic bias with the neural network reconstruction given that about the same amount of positive and negative bias is present across the 10 patients. The average absolute bias for DirectPET is 1.82\% with a maximum of 4.1\%. For DirectPET-50 the average is 2.04\% with a maximum of 4.60\% . This demonstrates that the neural networks trained on full and half-counts show similar low biases. Conversely, the OSEM+PSF-50 images have a negative bias around 50\%.       

The mean absolute error (MAE) is a common metric in deep learning research to indicate the accuracy of a trained model and is an explicit component of our loss function. Figure~\ref{quant}(b) shows the MAE for each patient image volume, again compared to the target OSEM+PSF volume. To prevent the many zero valued voxels present in the images from  skewing the metric, the absolute difference is only calculated for non-zero voxel values. For DirectPET, the resulting average MAE value across the 10 data sets is 33.07 Bq/ml. For DirectPET-50 and OSEM+PSF-50, the average MAE values are  33.57 Bq/ml and 265.7 Bq/ml, respectively. The nearly identical performance of the two neural networks is driven by MAE being a component of their loss function causing them to specifically optimize this measurement in the same way. 

Defined above, multi-scale structural similarity (MS-SSIM) values range from 0 to 1, where 0 means no similarity whatsoever while 1 means that the images are identical. Figure~\ref{quant}(d) plots MS-SSIM values for each reconstruction method with full count  OSEM+PSF serving as the reference. DirectPET and DirectPET-50 are seen to achieve similar values with both networks consistently at or above a value of 0.99 across the 10 data sets. The fact that MS-SSIM is included in the loss function is what once again leads to similar high performance. By contrast, OSEM+PSF-50 achieves an average structural similarity score of just 0.88.

\begin{figure*}[h]
	\centering
	\includegraphics[width=\textwidth]{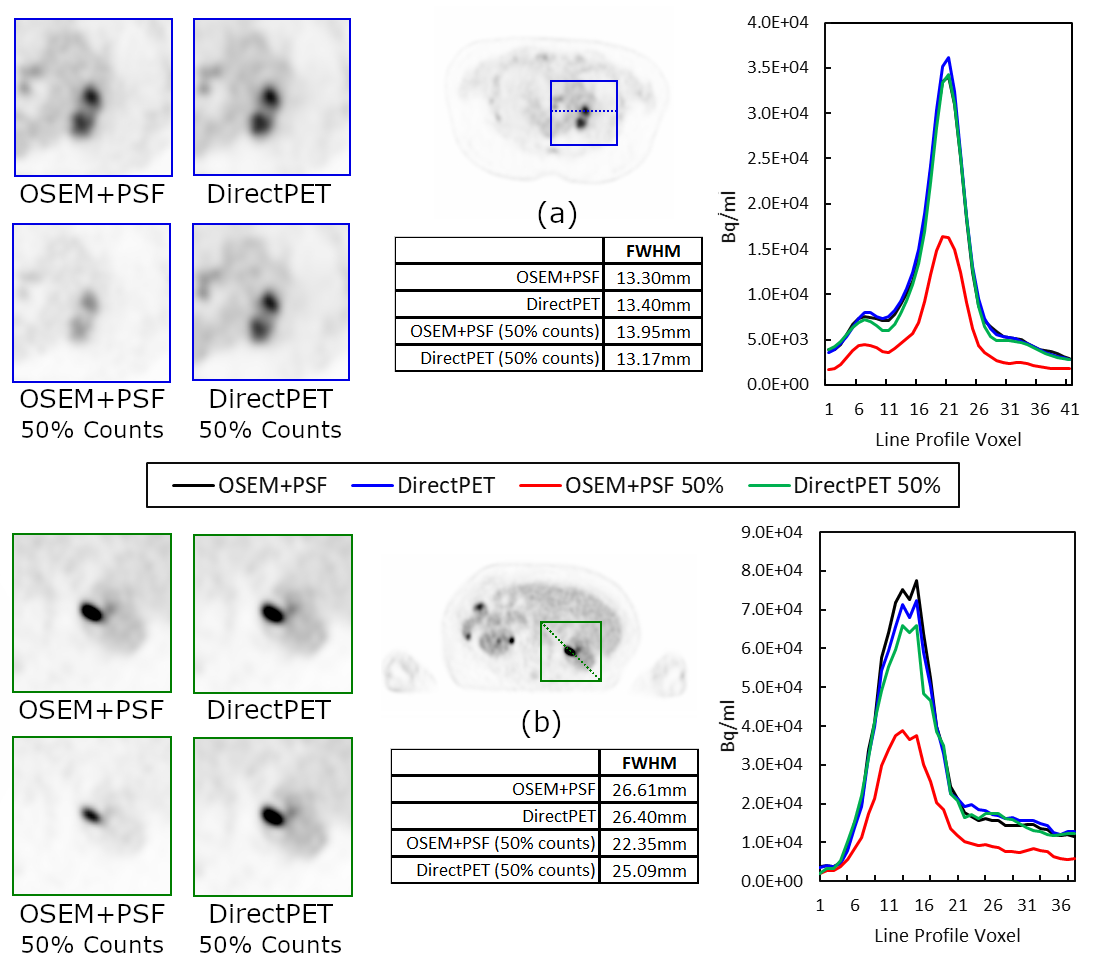}
	\caption{
	  Quantitative analysis of two lesions showing a line profile and full-width half-maximum (FWHM) for each reconstruction method.  
	}
	\label{line_profiles}
\end{figure*}

In Figure~\ref{line_profiles} we analyze two lesions in two independent images, and compare the benchmark OSEM+PSF reconstruction to DirectPET, DirectPET-50 and OSEM+PSF-50 reconstructions with line profiles, measures of full-width half-maximum (FWHM) and zoomed images. While the line profiles provide some intuition on the neural networks' performance on spatial resolution, the measurements are from patient data rather than a well defined point source or phantom. Spatial resolution can thus only be loosely inferred relative to the reference reconstruction. That is, if spatial resolution performance is poor, this should be evident in larger FWHM measurements of the neural network methods compared to the reference. In Figure~\ref{line_profiles}(a) the line profiles between both neural networks and the reference image are largely overlapping. The FWHM of DirectPET is 0.75\% larger than the reference, DirectPET-50 is 0.98\% smaller and OSEM+PSF-50 is 4.9\% larger. In Figure~\ref{line_profiles}(b) a larger lesion is analyzed and in this case the peak of the line profile is distinguishable between the two neural network methods and the reference, with the DirectPET neural network achieving a maximum value 6.7\% less than the reference and DirectPET-50 14.5\% less. 

From a qualitative aspect the line profiles, FWHM measurements and zoomed images seem to indicate reasonable neural network preservation of spatial resolution for this sample of lesions while also having a slightly higher SNR performance as discussed above. Although the focus of this paper is primarily on introducing direct neural network reconstruction as a viable method, the characterization of two randomly selected lesions can best be described as a preliminary and somewhat anecdotal study. We defer a quantitative examination of a large number of lesions along with the more well defined procedures of NEMA spatial resolution and image quality to future work.

\subsection{Qualitative Image Image Analysis}

\begin{figure*}[h]
	\centering
	\includegraphics[width=\textwidth]{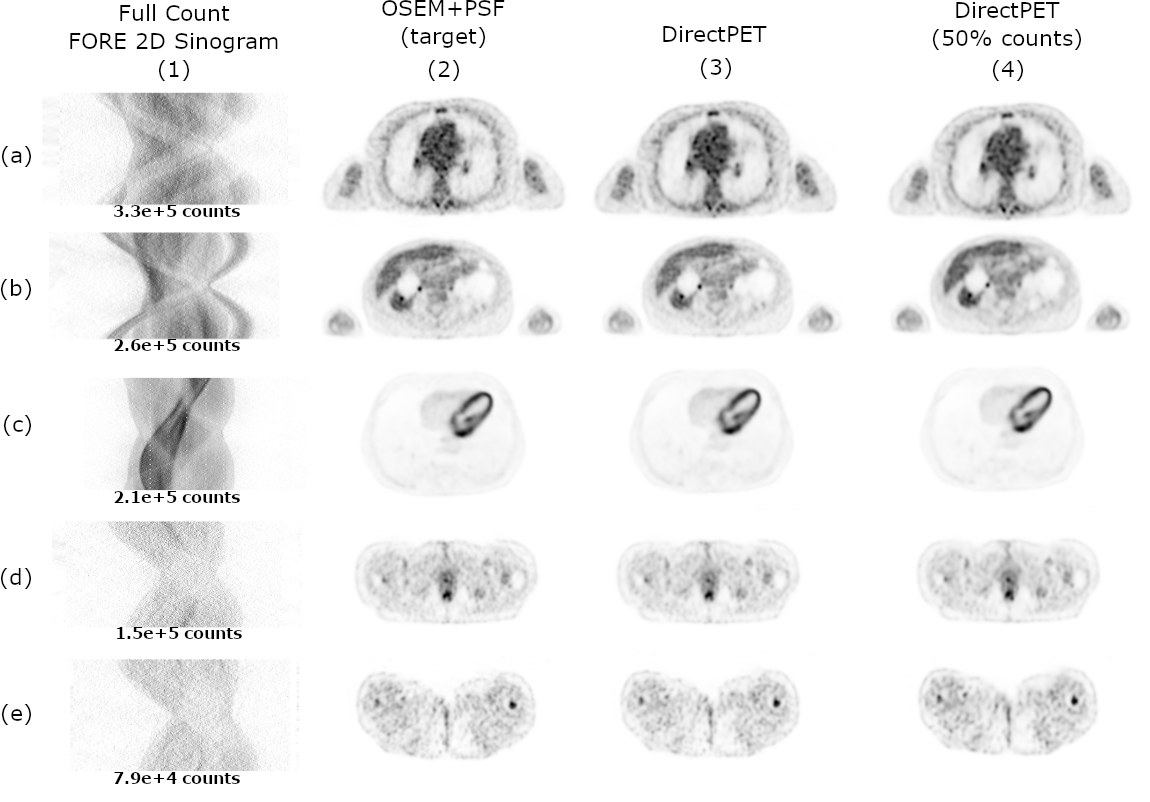}
	\caption{Full resolution test set reconstructions using OSEM+PSF, DirectPET and DirectPET-50 methods from a variety of body locations and count levels.
	}
	\label{image_compare}
\end{figure*}

In Figure~\ref{image_compare} we examine a sampling of 400x400 images slices from various regions of the body containing different count levels, ranging from 330k counts down to 79k counts. We again compare DirectPET and DirectPET-50 reconstructions to the OSEM+PSF reference.  Overall, visual comparison indicates strong similarity between the images. In particular,  areas of high tracer uptake, such as the chest in row (a), the lesions in row (b), and the heart in row (c), all show little difference to the reference images. Even the lower count images in rows (d) and (e) exhibit nearly identical areas of high uptake. On close inspection of areas of lower uptake, while still very similar to the reference, there are minor differences in intensity, structure and blurring present. Whether these subtle differences are clinically relevant is an open question to be explored in future research on neural network reconstruction to include lesion detectability, and observer studies.

\subsection{Limitations and Future Challenges}

One long term challenge to direct neural network reconstruction is understanding the boundaries and limits of a trained network when the quality and content of a medical image is at stake. The data driven nature of deep learning is both its most powerful strength and greatest weakness. As we and others have shown, a neural network can, somewhat counter intuitively, take a data set and essentially learn the mapping of physics and geometry of PET image reconstruction despite the inherent presence of significant noise. On the other hand, there is no mathematical or statistical guarantee that some unknown new data will be reconstructed with the same image quality. To combat this uncertainty large carefully curated data sets for training and validation could be maintained that uniformly represent the entire distribution of data a given neural network must learn (ethnicity, age, physical traits, tracer concentration, disease, etc.). While this someday may be possible, in the near term a more practical approach is to better understand the boundaries of the network's underlying learned distribution and develop techniques to classify or predict the reconstructed quality of previously unseen data, perhaps again utilizing a neural network for this task, and revert to conventional reconstruction techniques if the predicted image quality falls below a certain threshold.    

There is also the challenge of bootstrapping neural network reconstruction training when a new scanner geometry, radio-pharmaceutical or other aspect introduces the need to develop a fresh reconstruction (i.e. system matrix and corrections). At the outset of these new developments there is no data to train the neural network, and so skipping the step of developing conventional reconstruction methods, at the absolute least to create the data sets for neural network training, is not practical. It is perhaps possible and even likely to one day create data sets entirely from simulation that translate into high quality neural network images of patients in the physical world, but this still remains to be seen.   

A more logistical drawback of neural network reconstruction arises if a variety of reconstruction styles/parameters (attenuation, scatter, filter, noise, etc.) are desired. A single network is likely insufficient to meet this need, and the solution is to train multiple neural networks each specifically targeted to produce images with certain characteristics. Although this requires the training and management of multiple networks, the computational burden can possibly be mitigated through the use of transfer learning, which accelerates the training of a neural network on a new reconstruction by starting with a network previously trained on a similar reconstruction.

Regarding comparison with other algorithms, a limitation common to this work and all recent deep learning medical image reconstruction research is the lack of a robust platform to consistently compare results. Ideally, a large database containing raw data as well as associated state-of-the-art reconstructions would be publicly available for benchmarking new algorithms. The absence of such a database combined with the performance sensitivity of neural networks for a given data set in turn makes it challenging to directly compare the image quality of our method with existing image space, unrolled network and other direct reconstruction methods, thus limiting the comparison to differences in approach, computational efficiency and implementation complexity as discussed in the introduction.

\section{Conclusion}
We have presented DirectPET, a neural network capable of full size volume PET reconstruction directly from clinical sinogram data. The network contains three distinct segments, namely, one for encoding the sinogram data, one for converting the resulting data to image space, and one for refining and scaling the image for final presentation. The DirectPET network overcomes the key computational challenge of performing the domain transformation from sinogram to image space through the use of a novel Radon inversion layer enabling neural network reconstructions significantly larger (16x400x400) than any previous work. When batch operations are considered, the reconstruction of an entire PET whole-body scan (e.g., 400x400x400)is possible in a single, very fast forward pass of the network.  

Our work also shows the ability of a neural network to learn a higher quality reconstruction than the conventional PET benchmark of OSEM+PSF, if provided a training set with superior target images. This capability was demonstrated by removing half the counts in the raw data through list-mode thinning, training the DirectPET network to reconstruct full count images from half count sinogram data, and comparing the results to OSEM+PSF reconstructions on the decimated data. The results showed that the proposed neural network produced images nearly equivalent to using the full count data and superior to conventional reconstruction of the same data. While DirectPET was purposefully trained to match the performance of the standard Siemens OSEM+PSF reconstruction, similarly other suitable techniques such as maximum a posteriori (MAP) reconstruction or adding a non-local means filter, both of which are known to produce superior images, could have been used as the neural network training target. 

Looking toward future work, there are many possibilities in network architecture, loss functions and training optimization to explore, which will undoubtedly lead to more efficient reconstructions and even higher quality images. However, the biggest challenge with producing medical images is providing overall confidence on neural network reconstruction on unseen samples. While the understanding of deep learning techniques is growing and becoming less of a black box, future research should investigate the boundaries and limits of trained neural networks and how they relate to the underlying data distribution. Additionally, research to understand and quantify the clinical relevance and impact of neural network generated images will be an important step towards eventual adoption. 

\section{Acknowledgements}
The authors would like to thank Dr. Michael E. Casey for his advice and feedback,  Dr. Chuanyu Zhou for providing baseline OSEM+PSF reconstruction and Fourier rebinning software, advice and expertise, and The University of Tennessee Medical Center for providing PET whole-body patient data. 

\section{Disclosures}
No conflicts of interest, financial or otherwise, are declared by the authors. All data was acquired under the supervision of the institutional review board of The University of Tennessee Medical Center. 


\bibliography{report}   
\bibliographystyle{spiejour}   

\textbf{William Whiteley} is a Director of Software Engineering at Siemens Medical Solutions USA, Inc. working in molecular imaging and a PhD student at The University of Tennessee, Knoxville studying artificial intelligence and image reconstruction. He received his BSEE degree from Vanderbilt University, Nashville, TN and his MSEE degree from Stanford University, Stanford, CA. His research interests include artificial intelligence, deep learning, medical imaging, and data mining. 

\textbf{Wing K Luk} is a senior staff physicist with Siemens Medical Solutions USA, Inc. He received his BS degree in Engineering, MS degree in Physics and MS degree in Computer Science from University of Tennessee, Knoxville. His current research interests include PET image reconstruction, NUMA memory access optimization, dynamic imaging run-time load balancing and AI image reconstruction.

\textbf{Dr. Jens Gregor} received his PhD in Electrical Engineering from Aalborg University Denmark, in 1991. He then joined the faculty at the University of Tennessee, Knoxville, where he currently holds the position of Professor and Associate Department Head in the Department of Electrical Engineering and Computer Science. His research has covered different imaging modalities including X-ray and neutron CT, SPECT, and PET with applications ranging from industrial and security imaging to preclinical and clinical imaging. 

\newpage

\section{Figure Captions}

\paragraph{Figure 1.} The DirectPET imaging pipeline uses TOF Fourier rebinned PET data and X-ray CT based attenuation maps to generate PET image volumes.

\paragraph{Figure 2.} The DirectPET reconstruction neural network consists of three distinct segments each with a specific task: (a) the encoding segment is composed of convolutional layers that compress the sinogram input; (b) the domain transformation segment implements Radon inversion by applying masks to filter the compressed sinogram data into small fully connected networks for each of a number of image patches that are then combined to produce an initial image estimate; and (c) the refinement and scaling segment carries out denoising along with attenuation correction and applies super-resolution techniques to produce a final full-scale image.

\paragraph{Figure 3.} A single, fully-connected layer can be trained to learn the distinctive sinusoidal pattern associated with the Radon transform.

\paragraph{Figure 4.} The mask creation process begins with summing the raw pixel activation maps for an image patch and then undergoes a process of smoothing, morphological opening and closing, and thresholding to produce the final mask.

\paragraph{Figure 5.} The histograms in (a) and (b) show the relative distribution of slice counts in the Fourier rebinned sinograms for the training and test sets. (c) Shows the reconstruction time of a single FOV for both conventional methods and DirectPET demonstrating 7.2x and 4.9x improvement respectively.

\paragraph{Figure 6.} Quantitative measurements of the reference OSEM+PSF reconstructions, DirectPET, DirectPET trained on half-count input data and OSEM+PSF reconstructed with half-count input data show (a) the neural network methods produce images with less noise, (b) introduce a non-systematic bias of less than 5\%, (c) produce low absolute voxel error, and (d) produce images that are perceptually very similar to the target. These plots also show that the neural network trained with only half of the raw counts performed about the same as the network trained on full counts from a quantitative perspective.

\paragraph{Figure 7.} Quantitative analysis of two lesions showing a line profile and full-width half-maximum (FWHM) for each reconstruction method.

\paragraph{Figure 8.} Full resolution test set reconstructions using OSEM+PSF, DirectPET and DirectPET-50 methods from a variety of body locations and count levels.

\end{spacing}
\end{document}